\documentclass{article} 
\usepackage{nips15submit_e, times}
\pdfoutput=1
\usepackage[utf8]{inputenc}
\usepackage{hyperref}
\usepackage{url}
\usepackage{amsmath}
\usepackage{graphicx}
 
\hypersetup{colorlinks,%
            citecolor=black,%
            linkcolor=black,%
            urlcolor=black,%
}

\title{The Brain Uses Reliability of Stimulus Information when Making Perceptual Decisions}

\author{
Sebastian Bitzer\textsuperscript{1}\\
\texttt{sebastian.bitzer@tu-dresden.de} \\
\And
Stefan J.~Kiebel\textsuperscript{1}\\
\texttt{stefan.kiebel@tu-dresden.de}\\
\AND
\textnormal{\textsuperscript{1}Department of Psychology, Technische Universität Dresden, 01062 Dresden, Germany}
}

\nipsfinalcopy 

\begin{document}

\maketitle

\begin{abstract}
In simple perceptual decisions the brain has to identify a stimulus based on noisy sensory samples from the stimulus. Basic statistical considerations state that the reliability of the stimulus information, i.e., the amount of noise in the samples, should be taken into account when the decision is made. However, for perceptual decision making experiments it has been questioned whether the brain indeed uses the reliability for making decisions when confronted with unpredictable changes in stimulus reliability. We here show that even the basic drift diffusion model, which has frequently been used to explain experimental findings in perceptual decision making, implicitly relies on estimates of stimulus reliability. We then show that only those variants of the drift diffusion model which allow stimulus-specific reliabilities are consistent with neurophysiological findings. Our analysis suggests that the brain estimates the reliability of the stimulus on a short time scale of at most a few hundred milliseconds.
\end{abstract}

\section{Introduction}
In perceptual decision making participants have to identify a noisy stimulus. In typical experiments, only two possibilities are considered \cite{Gold2007}. The amount of noise on the stimulus is usually varied to manipulate task difficulty. With higher noise, participants' decisions are slower and less accurate. 

Early psychology research established that biased random walk models explain the response distributions (choice and reaction time) of perceptual decision making experiments \cite{John1967}. These models describe decision making as an accumulation of noisy evidence until a bound is reached and correspond, in discrete time, to sequential analysis \cite{Luce1986} as developed in statistics \cite{Wald1947a}. More recently, electrophysiological experiments provided additional support for such bounded accumulation models, see \cite{Gold2007} for a review.

There appears to be a general consensus that the brain implements the mechanisms required for bounded accumulation, although different models were proposed for how exactly this accumulation is employed by the brain \cite{Wang2002, Rao2004, Gold2007, Beck2008, Churchland2011}. An important assumption of all these models is that the brain provides the input to the accumulation, the so-called evidence, but the most established models actually do not define how this evidence is computed by the brain \cite{Luce1986, Wang2002, Bogacz2006, Gold2007}. In this contribution, we will show that addressing this question offers a new perspective on how exactly perceptual decision making may be performed by the brain.

Probabilistic models provide a precise definition of evidence: Evidence is the likelihood of a decision alternative under a noisy measurement where the likelihood is defined through a generative model of the measurements under the hypothesis that the considered decision alternative is true. In particular, this generative model implements assumptions about the expected distribution of measurements. Therefore, the likelihood of a measurement is large when measurements are assumed, by the decision maker, to be reliable and small otherwise. For modelling perceptual decision making experiments, the evidence input, which is assumed to be pre-computed by the brain, should similarly depend on the reliability of measurements as estimated by the brain. However, this has been disputed before, e.g.  \cite{Shadlen2008}. The argument is that typical experimental setups make the reliability of each trial unpredictable for the participant. Therefore, it was argued, the brain can have no correct estimate of the reliability. This issue has been addressed in a neurally inspired, probabilistic model based on probabilistic population codes (PPCs) \cite{Beck2008}. The authors have shown that PPCs can implement perceptual decision making without having to explicitly represent reliability in the decision process. This remarkable result has been obtained by making the comprehensible assumption that reliability has a multiplicative effect on the tuning curves of the neurons in the PPCs. Reliability, therefore, was implicitly represented in the tuning curves of model neurons which leaves the open question why tuning curves exhibit a multiplicative effect of reliability.

In this paper we will consider this question from a more conceptual perspective under the interpretation that the multiplicative effect on tuning curves results from adaptation of internal estimates of measurement reliability in the brain. We show that even a simple, widely used bounded accumulation model, the drift diffusion model, is based on some estimate of measurement reliability. Using this result, we will analyse the results of a perceptual decision making experiment \cite{Churchland2008} and will show that the recorded behaviour together with neurophysiological findings strongly favours the hypothesis that the brain weights evidence using a current estimate of measurement reliability, even when reliability changes unpredictably across trials.

This paper is organised as follows: We first introduce the notions of measurement, evidence and likelihood in the context of the experimentally well-established random dot motion (RDM) stimulus. We define these quantities formally by resorting to a simple probabilistic model which has been shown to be equivalent to the drift diffusion model \cite{Dayan2008a, Bitzer2014}. This, in turn, allows us to formulate three competing variants of the drift diffusion model that either do not use trial-dependent reliability (variant CONST), or do use trial-dependent reliability of measurements during decision making (variants DDM and DEPC, see below for definitions). Finally, using data of \cite{Churchland2008}, we show that only variants DDM and DEPC, which use trial-dependent reliability, are consistent with previous findings about perceptual decision making in the brain.

\section{Measurement, evidence and likelihood in the random dot motion stimulus}

The widely used random dot motion (RDM) stimulus consists of a set of randomly located dots shown within an invisible circle on a screen \cite{Newsome1988}. From one video frame to the next some of the dots move into one direction which is fixed within a trial of an experiment, i.e., a subset of the dots moves coherently in one direction. All other dots are randomly replaced within the circle. Although there are many variants of how exactly to present the dots \cite{Pilly2009}, the main idea is that the coherently moving dots indicate a motion direction which participants have to decide upon. By varying the proportion of dots which move coherently, also called the 'coherence' of the stimulus, the difficulty of the task can be varied effectively.

We will now consider what kind of evidence the brain can in principle extract from the RDM stimulus in a short time window, for example, from one video frame to the next, within a trial. For simplicity we call this time window 'time point' from here on, the idea being that evidence is accumulated over different time points, as postulated by bounded accumulation models in perceptual decision making \cite{Luce1986, Gold2007}. 

\begin{figure}[ht]
\begin{center}
\includegraphics[width=.8\linewidth]{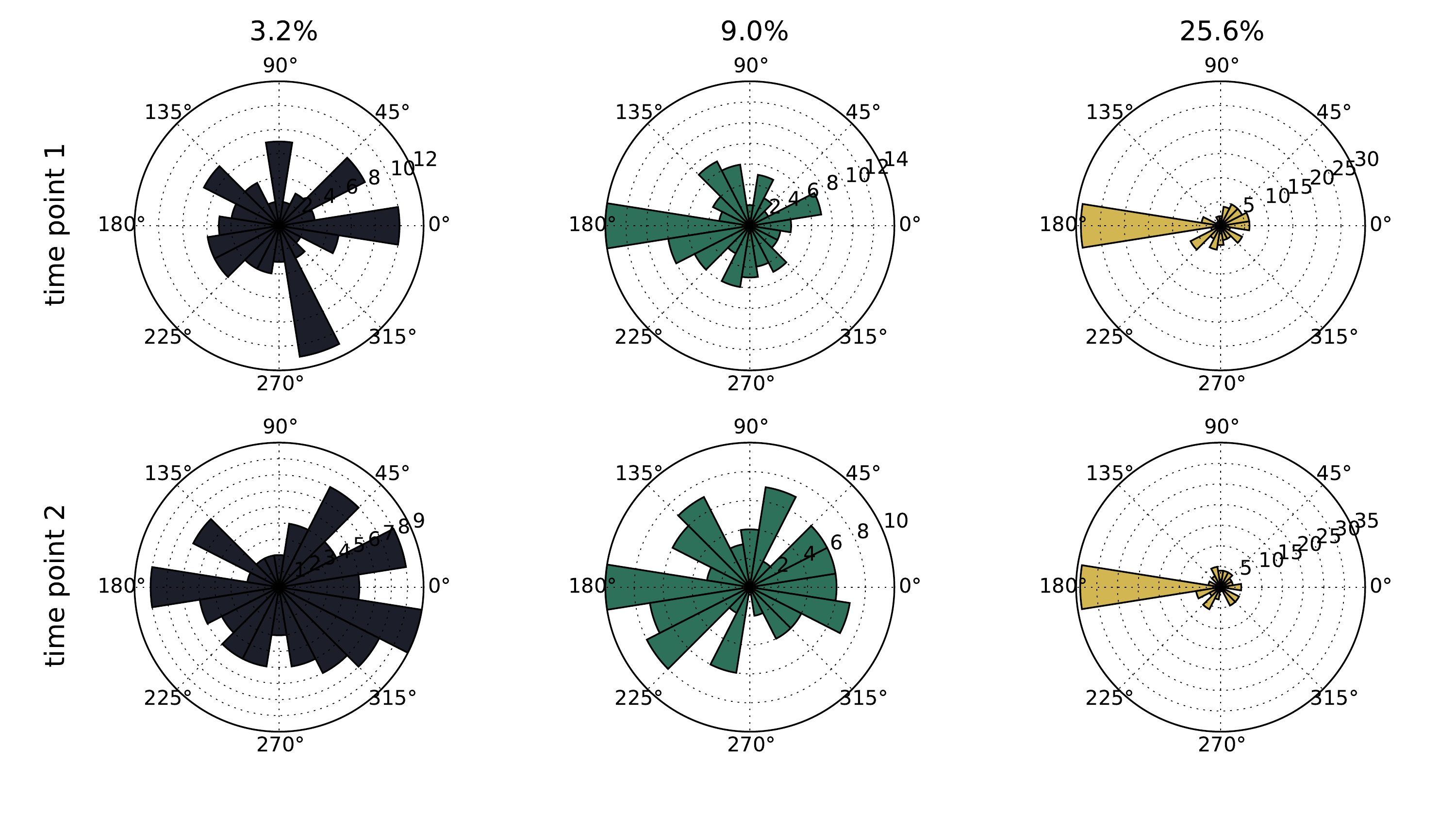}
\end{center}
\caption{Illustration of possible motion direction distributions that the brain can measure from an RDM stimulus. Rows are different time points, columns are different coherences. The true, underlying motion direction was 'left', i.e., $180^{\circ}$. For low coherence (e.g., 3.2\%) the measured distribution is very variable across time points and may indicate the presence of many different motion directions at any given time point. As coherence increases (from 9\% to 25.6\%), the true, underlying motion direction will increasingly dominate measured motion directions simultaneously leading to decreased variation of the measured distribution across time points.}
\label{fig:RDM}
\end{figure}

At a single time point, the brain can measure motion directions from the dots in the RDM display. By construction, a proportion of measurable motion directions will be into one specific direction, but, through the random relocation of other dots, the RDM display will also contain motion in random directions. Therefore, the brain observes a distribution of motion directions at each time point. This distribution can be considered a 'measurement' of the RDM stimulus made by the brain. Due to the randomness of each time frame, this distribution varies across time points and the variation in the distribution reduces for increasing coherences. We have illustrated this using rose histograms in Fig. \ref{fig:RDM} for three different coherence levels.

To compute the evidence for the decision whether the RDM stimulus contains predominantly motion to one of the two considered directions, e.g., left and right, the brain must check how strongly these directions are represented in the measured distribution, e.g., by estimating the proportion of motion towards left and right. We call these proportions evidence for left, $e_\mathrm{left}$, and evidence for right, $e_\mathrm{right}$. As the measured distribution over motion directions may vary strongly across time points, the computed evidences for each single time point may be unreliable. Probabilistic approaches weight evidence by its reliability such that unreliable evidence is not over-interpreted. The question is: Does the brain perform this reliability-based computation as well? More formally, for a given coherence, $c$, does the brain weight evidence by an estimate of reliability that depends on $c$: $l = e \cdot r(c)$\footnote{For convenience, we use imprecise denominations here. As will become clear below, $l$ is in our case a Gaussian log-likelihood, hence, the linear weighting of evidence by reliability.} and which we call 'likelihood', or does it ignore changing reliabilities and use a weighting unrelated to coherence: $e' = e \cdot \bar{r}$?

\section{Bounded accumulation models}
Bounded accumulation models postulate that decisions are made based on a decision variable. In particular, this decision variable is driven towards the correct alternative and is perturbed by noise. A decision is made, when the decision variable reaches a specific value. In the drift diffusion model, these three components are represented by drift, diffusion and bound \cite{Luce1986}. We will now relate the typical drift diffusion formalism to our notions of measurement, evidence and likelihood by linking the drift diffusion model to probabilistic formulations.

In the drift diffusion model, the decision variable evolves according to a simple Wiener process with drift. In discrete time the change in the decision variable $y$ can be written as
\begin{equation}\label{eq:ddm}
\delta y = y_{t} - y_{t-\delta t} = v\delta t + \sqrt{\delta t}s\epsilon_t
\end{equation}
where $v$ is the drift, $\epsilon_t \sim N(0, 1)$ is Gaussian noise and $s$ controls the amount of diffusion. This equation bears an interesting link to how the brain may compute the evidence. For example, it has been stated in the context of an experiment with RDM stimuli with two decision alternatives that the change in $y$, often called 'momentary evidence', "is thought to be a difference in firing rates of direction selective neurons with opposite direction preferences." \cite[Supp. Fig. 6]{Churchland2008} Formally:
\begin{equation}
\delta y = \rho_{\mathrm{left},t} - \rho_{\mathrm{right},t}
\end{equation}
where $\rho_{\mathrm{left},t}$ is the firing rate of the population selective to motion towards left at time point $t$. Because the firing rates $\rho$ depend on the considered decision alternative, they represent a form of evidence extracted from the stimulus measurement instead of the stimulus measurement itself (see our definitions in the previous section). It is unclear, however, whether the firing rates $\rho$ just represent the evidence ($\rho=e'$) or whether they represent the likelihood, $\rho = l$, i.e., the evidence weighted by coherence-dependent reliability.

To clarify the relation between firing rates $\rho$, evidence $e$ and likelihood $l$ we consider probabilistic models of perceptual decision making. Several variants have been suggested and related to other forms of decision making \cite{Rao2004, Yu2005a, Bogacz2006, Beck2008, Dayan2008a, Solway2012, Huang2012}. For its simplicity, which is sufficient for our argument, we here consider the model presented in \cite{Bitzer2014} for which a direct transformation from probabilistic model to the drift diffusion model has already been shown. This model defines two Gaussian generative models of measurements which are derived from the stimulus:
\begin{equation}
p(x_t| \mathrm{left}) = N(-1, \delta t\hat{\sigma}^2) \qquad p(x_t| \mathrm{right}) = N(1, \delta t\hat{\sigma}^2)
\end{equation}
where $\hat{\sigma}$ represents the variability of measurements expected by the brain. Similarly, it is assumed that the measurements $x_t$ are sampled from a Gaussian with variance $\sigma^2$ which captures variance both from the stimulus and due to other noise sources in the brain:
\begin{equation}
x_t \sim N(\pm 1, \delta t \sigma^2).
\end{equation}
Evidence for a decision is computed in this model by calculating the likelihood of a measurement $x_t$ under the hypothesised generative models. To be precise we consider the log-likelihood which is
\begin{equation}\label{eq:likelihood}
l = -\log(\sqrt{2\pi\delta t}\hat{\sigma}) - \frac{1}{2}\frac{(x_t \pm 1)^2}{\delta t\hat{\sigma}^2}.
\end{equation}
There are three important points: 1) The first term on the right hand side means that $l$ increases independently of the stimulus $x_t$ for decreasing $\hat{\sigma}$. This contribution cancels when the difference between the likelihoods for left and right is computed. 2) The likelihood is large for a measurement $x_t$, when $x_t$ is close to the values hypothesised for the decision alternatives, i.e., $-1$ and $1$. 3) The contribution of the stimulus is weighted by the assumed reliability $r = \hat{\sigma}^{-2}$.

This model of the RDM stimulus is simple but captures the most important properties of the stimulus. In particular, a high coherence RDM stimulus has a large proportion of motion in the correct direction with very low variability of measurements whereas a low coherence RDM stimulus tends to have lower proportions of motion in the correct direction, with high variability (cf. Fig. \ref{fig:RDM}). The Gaussian model captures these properties by adjusting the noise variance such that a high coherence corresponds to low noise and low coherence to high noise: Under high noise the values $x_t$ will vary strongly and tend to be rather distant from $-1$ and $1$, whereas for low noise the values $x_t$ will be close to $-1$ or $1$ with low variability. Hence, as expected, the model produces large evidences/likelihoods for low noise and small evidences/likelihoods for high noise.

This intuitive relation between stimulus and probabilistic model is the basis for us to proceed to show that the reliability of the stimulus $r$, connected to the coherence level $c$, appears at a prominent position in the drift diffusion model. Crucially, the drift diffusion model can be derived as the sum of log-likelihood ratios across time \cite{Luce1986, Bogacz2006, Dayan2008a, Bitzer2014}. In particular, a discrete time drift diffusion process can be derived by subtracting the likelihoods of Eq. \eqref{eq:likelihood}:
\begin{equation}
\delta y = l_\mathrm{right} - l_\mathrm{left} = \frac{(x_t + 1)^2 - (x_t - 1)^2}{2\delta t\hat{\sigma}^2} = \frac{2rx_t}{\delta t}.
\end{equation}
Consequently, the change in $y$ is Gaussian: $\delta y \sim N(2r/\delta t, 4r^2\sigma^2/\delta t)$. This replicates the model described in \cite[Supp. Fig. 6]{Churchland2008} where the parameterisation of the model, however, more directly followed that of the Gaussian distribution and did not explicitly take time into account: $\delta y \sim N(Kc, S^2)$, where $K$ and $S$ are free parameters and $c$ is coherence of the RDM stimulus. By analogy to the probabilistic model, we, therefore, see that the model in \cite{Churchland2008} implicitly assumes that reliability $r$ depends on coherence $c$.

More generally, the parameters of the drift diffusion model of Eq. \eqref{eq:ddm} and that of the probabilistic model can be expressed as functions of each other \cite{Bitzer2014}:
\begin{equation}\label{eq:drift}
v = \pm \frac{2}{\delta t^2 \hat{\sigma}^2} = \pm r \frac{2}{\delta t^2}
\end{equation}
\begin{equation}\label{eq:diffusion}
s = \frac{2\sigma}{\delta t \hat{\sigma}^2} = r \frac{2\sigma}{\delta t}.
\end{equation}
These equations state that both drift $v$ and diffusion $s$ depend on the assumed reliability $r$ of the measurements $x$. Does the brain use and necessarily compute this reliability which depends on coherence? In the following section we answer this question by comparing how well three variants of the drift diffusion model, that implement different assumptions about $r$, conform to experimental findings.

\section{Use of reliability in perceptual decision making: experimental evidence}
We first show that different assumptions about the reliability $r$ translate to variants of the drift diffusion model. We then fit all variants to behavioural data (performances and mean reaction times) of an experiment for which neurophysiological data has also been reported \cite{Churchland2008} and demonstrate that only those variants which allow reliability to depend on coherence level lead to accumulation mechanisms which are consistent with the neurophysiological findings.

\subsection{Drift diffusion model variants}
For the drift diffusion model of Eq. \eqref{eq:ddm} the accuracy $A$ and mean decision time $T$ predicted by the model can be determined analytically \cite{Bogacz2006}:
\begin{equation}\label{eq:accuracy}
A = 1 - \frac{1}{1 + \exp(\frac{2vb}{s^2})}
\end{equation}
\begin{equation}\label{eq:meanRT}
T = \frac{b}{v} \tanh\left( \frac{vb}{s^2} \right)
\end{equation}
where $b$ is the bound. These equations highlight an important caveat of the drift diffusion model: Only two of the three parameters can be determined uniquely from behavioural data. For fitting the model one of the parameters needs to be fixed. In most cases, the diffusion $s$ is set to $c=0.1$ arbitrarily \cite{Bogacz2006}, or is fit with a constant value across stimulus strengths \cite{Churchland2008}. We call this standard variant of the drift diffusion model the DDM.

If $s$ is constant across stimulus strengths, the other two parameters of the model must explain differences in behaviour, between stimulus strengths, by taking on values that depend on stimulus strength. Indeed, it has been found that primarily drift $v$ explains such differences, see also below. Eq. \eqref{eq:drift} states that drift depends on estimated reliability $r$. So, if drift varies across stimulus strengths, this strongly suggests that $r$ must vary across stimulus strengths, i.e., that $r$ must depend on coherence: $r(c)$. However, the drift diffusion formalism allows for two other obvious variants of parameterisation. One in which the bound $b$ is constant across stimulus strengths, $b = \bar{b}$, and, conversely, one in which drift $v$ is constant across stimulus strengths, $v = \bar{v} \propto \bar{r}$ (Eq. \ref{eq:drift}). We call these variants DEPC and CONST, respectively, for their property to weight evidence by reliability that either depends on coherence, $r(c)$, or not, $\bar{r}$.

\subsection{Experimental data}
In the following we will analyse the data presented in \cite{Churchland2008}. This data set has two major advantages for our purposes: 1) Reported accuracies and mean reaction times (Fig. 1d,f) are averages based on 15,937 trials in total. Therefore, noise in this data set is minimal (cf. small error bars in Fig. 1d,f) such that any potential effects of overfitting on found parameter values will be small, especially in relation to the effect induced by different stimulus strengths. 2) The behavioural data is accompanied by recordings of neurons which have been implicated in the decision making process. We can, therefore, compare the accumulation mechanisms resulting from the fit to behaviour with the actual neurophysiological recordings. Furthermore, the structure of the experiments was such that the stimulus in subsequent trials had random strength, i.e., the brain could not have estimated stimulus strength of a trial before the trial started.

In the experiment of \cite{Churchland2008}, that we consider here, two monkeys performed a two-alternative forced choice task based on the RDM stimulus. Data for eight different coherences were reported. To avoid ceiling effects, which prevent the unique identification of parameter values in the drift diffusion model, we exclude those coherences which lead to an accuracy of 0.5 (random choices) or to an accuracy of 1 (perfect choices). The behavioural data of the remaining six coherence levels are presented in Table \ref{tab:behaviour}.

\begin{table}[ht]
\caption{Behavioural data of \cite{Churchland2008} used in our analysis. RT = reaction time.}
\begin{center}
\begin{tabular}{rccccc}
coherence (\%): & 3.2 & 6.4 & 9 & 12 & 25.6\\
accuracy (fraction): & 0.63 & 0.76 & 0.79 & 0.89 & 0.99\\
mean RT (ms): & 613 & 590 & 580 & 535 & 440\\
\end{tabular}
\end{center}
\label{tab:behaviour}
\end{table}

The analysis of \cite{Churchland2008} revealed a nondecision time, i.e., a component of the reaction time that is unrelated to the decision process (cf. \cite{Luce1986}) of ca. 200ms. Using this estimate, we determined the mean decision time $T$ by subtracting 200ms from the mean reaction times shown in Table \ref{tab:behaviour}.

The main findings for the neural recordings, which replicated previous findings \cite{Roitman2002, Gold2007}, were that i) firing rates at the end of decisions were similar and, particularly, showed no significant relation to coherence \cite[Fig. 5]{Churchland2008} whereas ii) the buildup rate of neural firing within a trial had an approximately linear relation to coherence \cite[Fig. 4]{Churchland2008}.

\subsection{Fits of drift diffusion model variants to behaviour}
We can easily fit the model variants (DDM, DEPC and CONST) to accuracy $A$ and mean decision time $T$ using Eqs. \eqref{eq:accuracy} and \eqref{eq:meanRT}. In accordance with previous approaches we selected values for the respective redundant parameters. Since the redundant parameter value, or its inverse, simply scales the fitted parameter values (cf. Eqs. \ref{eq:accuracy} and \ref{eq:meanRT}), the exact value is irrelevant and we fix, in each model variant, the redundant parameter to 1.

\begin{figure}[ht]
\begin{center}
\includegraphics[width=\linewidth]{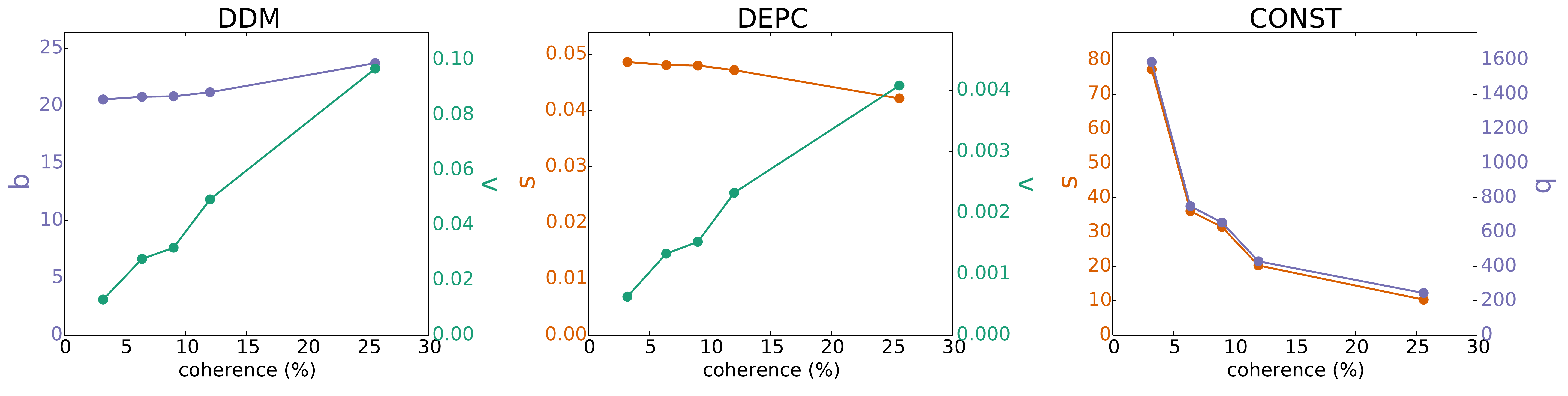}
\end{center}
\caption{Fitting results: values of the free parameters, that replicate the accuracy and mean RT recorded in the experiment (Table \ref{tab:behaviour}), in relation to coherence. The remaining, non-free parameter was fixed to 1 for each variant. Left: the DDM variant with free parameters drift $v$ (green) and bound $b$ (purple). Middle: the DEPC variant with free parameters $v$ and diffusion $s$ (orange). Right: the CONST variant with free parameters $s$ and $b$.}
\label{fig:fittedpars}
\end{figure}

Fig. \ref{fig:fittedpars} shows the inferred parameter values. In congruence with previous findings, the DDM variant explained variation in behaviour due to an increasing coherence mostly with an increasing drift $v$ (green in Fig. \ref{fig:fittedpars}). Specifically, drift and coherence appear to have a straightforward, linear relation. The same finding holds for the DEPC variant. In contrast to the DDM variant, however, which also exhibited a slight increase in the bound $b$ (purple in Fig. \ref{fig:fittedpars}) with increasing coherence, the DEPC variant explained the corresponding differences in behaviour by decreasing diffusion $s$ (orange in Fig. \ref{fig:fittedpars}). As the drift $v$ was fixed in CONST, this variant explained coherence-dependent behaviour with large and almost identical changes in both diffusion $s$ and bound $b$ such that large parameter values occurred for small coherences and the relation between parameters and coherence appeared to be quadratic.

\begin{figure}[ht]
\begin{center}
\includegraphics[width=\linewidth]{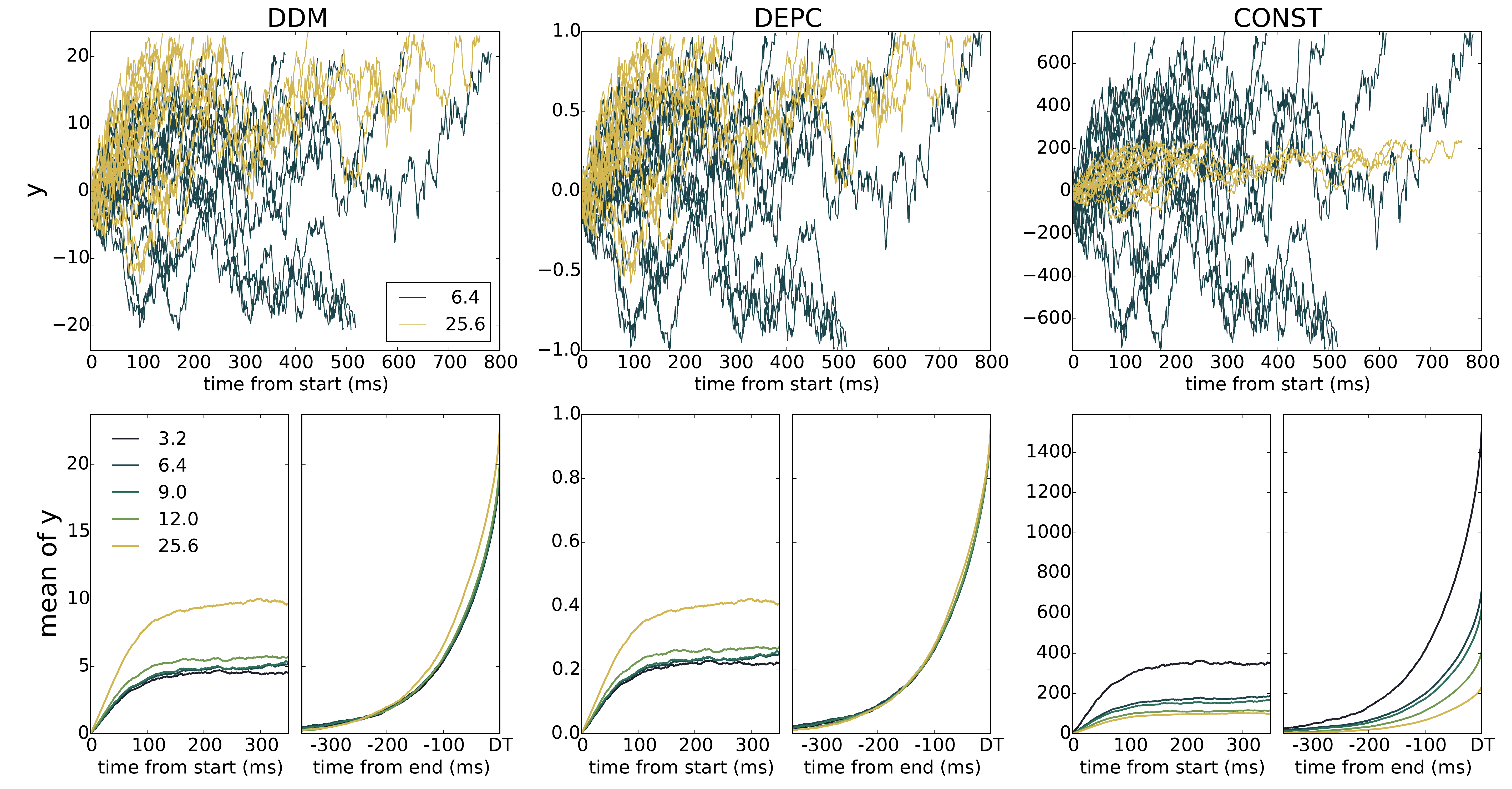}
\end{center}
\caption{Drift-diffusion properties of fitted model variants. Top row: 15 example trajectories of $y$ for different model variants with fitted parameters for 6.4\% (blue) and 25.6\% (yellow) coherence. Trajectories end when they reach the bound for the first time which corresponds to the decision time in that simulated trial. Notice that the same random samples of $\epsilon$ were used across variants and coherences. Bottom row: Trajectories of $y$ averaged over trials in which the first alternative (top bound) was chosen for the three model variants. Format of the plots follows that of \cite[Supp. Fig. 4]{Churchland2011}: Left panels show the buildup of $y$ from the start of decision making for the 5 different coherences. Right panels show the averaged drift diffusion trajectories when aligned to the time that a decision was made.}
\label{fig:exmptraj}
\end{figure}

We further investigated the properties of the model variants with the fitted parameter values. The top row of Fig. \ref{fig:exmptraj} shows example drift diffusion trajectories ($y$ in Eq. \eqref{eq:ddm}) simulated at a resolution of 1ms for two coherences. Following \cite{Churchland2008}, we interpret $y$ as the decision variables represented by the firing rates of neurons in monkey area LIP. These plots exemplify that the DDM and DEPC variants lead to qualitatively very similar predictions of neural responses whereas the trajectories produced by the CONST variant stand out, because the neural responses to large coherences are predicted to be smaller than those to small coherences. 

We have summarised predicted neural responses to all coherences in the bottom row of Fig. \ref{fig:exmptraj} where we show averages of $y$ across 5000 trials either aligned to the start of decision making (left panels) or aligned to the decision time (right panels). These plots illustrate that the DDM and DEPC variants replicate the main neurophysiological findings of \cite{Churchland2008}: Neural responses at the end of the decision were similar and independent of coherence. For the DEPC variant this was built into the model, because the bound was fixed. For the DDM variant the bound shows a small dependence on coherence, but the neural responses aligned to decision time were still very similar across coherences. The DDM and DEPC variants, further, replicate the finding that the buildup of neural firing depends approximately linear on coherence (normalised mean square error of a corresponding linear model was 0.04 and 0.03, respectively). In contrast, the CONST variant exhibited an inverse relation between coherence and buildup of predicted neural response, i.e., buildup was larger for small coherences. Furthermore, neural responses at decision time strongly depended on coherence. Therefore, the CONST variant, as the only variant which does not use coherence-dependent reliability, is also the only variant which is clearly inconsistent with the neurophysiological findings.

\section{Discussion}
We have investigated whether the brain uses online estimates of stimulus reliability when making simple perceptual decisions. From a probabilistic perspective fundamental considerations suggest that using accurate estimates of stimulus reliability lead to better decisions, but in the field of perceptual decision making it has been questioned that the brain estimates stimulus reliability on the very short time scale of a few hundred milliseconds. By using a probabilistic formulation of the most widely accepted model we were able to show that only those variants of the model which assume online reliability estimation are consistent with reported experimental findings.

Our argument is based on a strict distinction between measurements, evidence and likelihood which may be briefly summarised as follows: Measurements are raw stimulus features that do not relate to the decision, evidence is a transformation of measurements into a decision relevant space reflecting the decision alternatives and likelihood is evidence scaled by a current estimate of measurement reliabilities. It is easy to overlook this distinction at the level of bounded accumulation models, such as the drift diffusion model, because these models assume a pre-computed form of evidence as input. However, this evidence has to be computed by the brain, as we have demonstrated based on the example of the RDM stimulus and using behavioural data.

We chose one particular, simple probabilistic model, because this model has a direct equivalence with the drift diffusion model which was used to explain the data of \cite{Churchland2008} before. Other models may have not allowed conclusions about reliability estimates in the brain. In particular, \cite{Bitzer2014} introduced an alternative model that also leads to equivalence with the drift diffusion model, but explains differences in behaviour by different mean measurements and their representations in the generative model. Instead of varying reliability across coherences, this model would vary the difference of means in the second summand of Eq. \eqref{eq:likelihood} directly without leading to any difference on the drift diffusion trajectories represented by $y$ of Eq. \eqref{eq:ddm} when compared to those of the probabilistic model chosen here. The interpretation of the alternative model of \cite{Bitzer2014}, however, is far removed from basic assumptions about the RDM stimulus: Whereas the alternative model assumes that the reliability of the stimulus is fixed across coherences, the noise in the RDM stimulus clearly depends on coherence. We, therefore, discarded the alternative model here.

As a slight caveat, the neurophysiological findings, on which we based our conclusion, could have been the result of a search for neurons that exhibit the properties of the conventional drift diffusion model (the DDM variant). We cannot exclude this possibility completely, but given the wide range and persistence of consistent evidence for the standard bounded accumulation theory of decision making \cite{Gold2007, Hanks2015} we find it rather unlikely that the results in \cite{Roitman2002} and \cite{Churchland2008} were purely found by chance. Even if our conclusion about the rapid estimation of reliability by the brain does not endure, our formal contribution holds: We clarified that the drift diffusion model in its most common variant (DDM) is consistent with, and even implicitly relies on, coherence-dependent estimates of measurement reliability.

In the experiment of \cite{Churchland2008} coherences of the RDM stimulus were chosen randomly for each trial. Consequently, participants could not predict the reliability of the RDM stimulus for the upcoming trial, i.e., the participants' brains could not have had a good estimate of stimulus reliability at the start of a trial. Yet, our analysis strongly suggests that coherence-dependent reliabilities were used during decision making. The brain, therefore, must had adapted reliability within trials even on the short timescale of a few hundred milliseconds. On the level of analysis dictated by the drift diffusion model we cannot observe this adaptation. It only manifests itself as a change in mean drift that is assumed to be constant within a trial. First models of simultaneous decision making and reliability estimation have been suggested \cite{Deneve2012}, but clearly more work in this direction is needed to elucidate the underlying mechanism used by the brain. 

\bibliography{reliability}

\end{document}